\documentclass[12pt]{article}
\usepackage{amsmath}
\usepackage{amssymb}
\usepackage{graphicx,bbm,mathrsfs}
\usepackage{color}

\newcommand{\lsim}{\raisebox{-0.13cm}{~\shortstack{$<$ \\[-0.07cm] $\sim$}}~} 
\newcommand{\gsim}{\raisebox{-0.13cm}{~\shortstack{$>$ \\[-0.07cm] $\sim$}}~} 
\newcommand{\beq}{\begin{eqnarray}} 
\newcommand{\eeq}{\end{eqnarray}} 
\newcommand{\tb}{\tan\beta} 
 
\usepackage{listings}
\usepackage{caption}
\usepackage{cite}

\begin{document}

{\small \hfill CERN-PH-TH/2016--109, LPT-Orsay--16--41, MAN/HEP/2016/07}

\vspace*{1mm}

\begin{center} 
\mbox{\Large{\bf Enhanced Rates for Diphoton Resonances in the MSSM}}\\[4mm]
{\sc Abdelhak~Djouadi$^{1,2}$} and {\sc Apostolos Pilaftsis$^{2,3}$ }\\[4mm]
{\small 
$^1$ Laboratoire de Physique Th\'eorique,  CNRS and Universit\'e Paris-Sud, \\  
B\^at. 210, F--91405 Orsay Cedex, France \\[1mm]
$^2$ Theory Department, CERN, CH 1211 Geneva 23, Switzerland \\[1mm]
$^3$ Consortium for Fundamental Physics, School of Physics and Astronomy, \\
University of Manchester, Manchester, M13 9PL, United Kingdom
}
\end{center}


\begin{abstract}
We propose a simple mechanism for copiously producing heavy Higgs bosons with enhanced 
decay rates to two photons at the LHC, within the context of the Minimal Supersymmetric extension of the Standard Model (MSSM). In the CP--conserving limit of the theory, such a diphoton resonance may be identified with the heavier CP--even $H$ boson, whose gluon--fusion production and decay into two photons are enhanced by loops of the lightest supersymmetric partner of the top quark $\tilde{t}_1$ when its mass $m_{\tilde{t}_1}$ happens to be near the $\tilde{t}^*_1\tilde{t}_1$ threshold, i.e.~for $m_{\tilde{t}_1}\!\simeq \!\frac12 M_H$. The scenario requires a relatively low supersymmetry-breaking scale~$M_S\lsim 1$~TeV, but large values of the higgsino mass parameter, $\mu \gsim 1$~TeV, that lead to a strong $H \tilde{t}^*_1 \tilde{t}_1$ coupling. Such parameters can accommodate the observed mass and standard--like couplings of the 125~GeV $h$ boson in the MSSM, while satisfying all other constraints from the LHC and dark matter searches. Additional enhancement to the diphoton rate could be provided by Coulombic QCD corrections and, to a lesser extent, by resonant contributions due to $\tilde{t}_1^* \tilde{t}_1$ bound states. To discuss the characteristic features of such a scenario, we consider as an illustrative example the case of a diphoton resonance with a mass of approximately 750 GeV, for which an excess was observed in the early LHC 13 TeV data and which later turned out to be simply a statistical fluctuation. 

\end{abstract}

\vspace*{3mm}

In December 2015, the ATLAS and CMS collaborations have reported an excess in the 13 TeV LHC data corresponding to a possible resonance $\Phi$ with a mass of approximately 750~GeV decaying into two photons~\cite{diphoton}. With the collection of more data in 2016, this initial diphoton excess turned out to be simply a statistical fluctuation and faded away~\cite{LHC2}.  In the meantime, a large number of phenomenological papers were written~\cite{phi} interpreting 
the excess in terms of a resonance and attempting to explain the very large initial diphoton rate. Indeed, assuming that the new state~$\Phi$ is a scalar boson, the production cross section in gluon--fusion $\sigma (gg \to \Phi )$ times the two--photon decay branching ratio ${\rm BR}(\Phi \to \gamma\gamma )$ was reported to be of order of a few femtobarns and such rates were very difficult to accommodate in minimal and theoretically well-motivated scenarios beyond the Standard Model~(SM)~\cite{phi}. As we need to stay alert for such unexpected surprises of New Physics at future LHC runs, the study of new mechanisms that lead to enhanced production rates for such diphoton resonances remains an interesting topic on its own right. In this paper, we consider diphoton resonances in one such scenario: the Minimal Supersymmetric extension of the SM~(MSSM)~\cite{MSSM,Review}, softly broken at scales $M_S = {\cal O}($1~TeV) for phenomenological reasons.  We investigate a few possibilities that lead to a large enhancement of the $pp\to \gamma\gamma$ rate which, for instance, could have explained the too large 750 GeV excess in the initial LHC 13 TeV data in terms of New Physics.

In the MSSM, two Higgs doublets are needed to break the electroweak symmetry leading to three neutral and two charged physical states. The $\Phi$ resonance could have corresponded to either the heavier CP--even $H$ or the CP--odd $A$ bosons \cite{ADM}, both contributions of which may be added individually at the cross-section level. The heavy neutral $H$ and $A$ bosons are degenerate in mass $M_H \approx M_A$ in the so--called decoupling regime $M_A\! \gg \!M_Z$ in which the lighter CP--even $h$ state, corresponding to the observed 125 GeV Higgs boson, has SM--like couplings as indicated by the LHC data \cite{PDG}. Nevertheless, it has been shown~\cite{ADM} that in most of the MSSM parameter space, a diphoton rate of ${\cal O}$(a few$~{\rm fb})$ cannot be generated using purely the MSSM particle content. Indeed, although the $\Phi=H/A$ Yukawa couplings to top quarks are sizeable for small values of the well--known ratio $\tb$ of the two--Higgs--doublet vacuum expectation values, the only input besides $M_A$ that is needed to characterize the MSSM Higgs sector (even when the important radiative corrections are taken into account \cite{hMSSM}), the top quark cannot generate sizeable enough loop contributions to the $gg\to H/A$ and $H/A\to \gamma\gamma$ processes to accommodate such a diphoton rate.  The supersymmetric particles give in general too small loop contributions because their couplings to the Higgs bosons are not proportional to the masses and decouple like $\propto 1/ M_S^2$ for large enough sparticle masses.

In this Letter we show that there exists a small but vital area of parameter space, in which large production rates of ${\cal O}(1~{\rm fb})$ for diphoton resonances at the LHC with $\sqrt s=13$--14 TeV can be accounted for, entirely {\em within} the restricted framework of the MSSM. In the CP-conserving limit of the theory,  the CP-even $H$ boson of the MSSM would be produced through the effective $Hgg$ and $H\gamma \gamma$ couplings, which are enhanced via loops involving the lightest top squark~$\tilde t_1$. The state $\tilde t_1$ will have significant loop contribution if its mass $m_{\tilde{t}_1}$ happens to be near the $\tilde{t}^*_1\tilde{t}_1$ threshold, $m_{\tilde{t}_1}  \simeq \frac12 M_H$. Given that stoponium $\Sigma_{\tilde{t}} \equiv (\tilde{t}^*_1\tilde{t}_1)$ bound states can be formed in this kinematic region \cite{Stoponium}, the diphoton rate will be further enhanced by resonant contributions to the amplitude thanks to the~$\Sigma_{\tilde{t}}$ states.  In addition, assuming a Higgs mass 
$M_H \lsim 1$ TeV, large values of the  Higgsino mass parameter $\mu$  are required, e.g.~$\mu\!\gsim$1~TeV, for a stop SUSY-breaking scale $M_S\!\sim \frac12$--1~TeV.  Such values enhance the strength of the $H\tilde{t}^*_1\tilde{t}_1$ coupling, through the so-called $F$-term contribution from the Higgs doublet superfield~$H_u$ that couples to up-type quark superfields.  Another smaller source of enhancement arises from the stop mixing parameter $A_t$, which can still play a significant role if the ratio $\tan\beta$ is relatively low,~i.e.~for $\tan\beta \lsim 10$.  

Besides comfortably allowing ${\cal O}(1~{\rm fb})$ diphoton rate, such parameter scenarios can naturally describe the observed SM-like $h$ state with a mass of 125~GeV, for $\tan\beta \gsim 5$ (after allowing for all theoretical uncertainties of a few GeV due to higher order effects), and comply with all present constraints on the supersymmetric particle spectrum \cite{PDG}. Here, we assume that the top squark $\tilde t_1$ is the lightest or next-to-lightest {\em visible} supersymmetric particle, for which a lower-mass gravitino or a bino nearly degenerate with~$\tilde t_1$ can successfully play the role of the dark matter in the Universe, respectively.

For illustration, let us now discuss in detail an example in which the diphoton resonance $\Phi$ is the one that could have corresponded to the excess observed in early 13 TeV data and how it could have been explained in the MSSM. The $\Phi$ state may be either the CP-even $H$ boson or the CP--odd $A$ scalar which, in the decoupling limit, have both suppressed couplings to~$W^\pm$ and $Z$ gauge bosons, and similar couplings to fermions. The latter are controlled by $\tan\beta$, with $1\lsim \tb \lsim 60$. For values $\tan\beta \lsim 5$, the only important Yukawa coupling~$\lambda_{\Phi ff}$ is the one of the top quark, while for $\tb \gsim 10$, the couplings to bottom quarks and $\tau$ leptons are enhanced, i.e. $\lambda_{\Phi ff}= \sqrt 2 m_f/v \times \hat g_{\Phi ff }$ with $\hat g_{\Phi tt }= \cot\beta$ and $\hat g_{\Phi bb } = \hat g_{\Phi \tau\tau } = \tan\beta$ at the tree level.  Nevertheless, for a mass $M_\Phi \! \approx \! 750$ GeV, values $\tb \gsim 20$ are excluded by the search of $A/H \to \tau\tau$ resonances \cite{LHC-tautau}, while $\tb$ values not too close to unity can be accommodated by the search for $H/A \to t\bar t$ resonance~\cite{LHC-tt}.

At the LHC, the $\Phi = H/A$ states are mainly produced in the $gg\to \Phi$ fusion mechanism that is mediated by a $t$-quark loop with cross sections at $\sqrt s \!=\! 13$ TeV of about $\sigma (gg \to A) \! \approx \! 1.3$ pb and $\sigma(gg\to H)=0.8$ pb for $M_\Phi \! \approx \! 750$ GeV and $\tb \! \approx \! 1$ \cite{LHC-prod}.  The $H/A$ states will then mainly decay into top quark pairs with partial ($\approx$ total) widths that are of order $\Gamma_\Phi \approx 30$ GeV. The two-photon decays of the $H$ and $A$ states are generated by the top quark loop only, and the branching ratio for the relevant inputs are: BR$(A \to \gamma\gamma) \approx 7 \times 10^{-6}$ and BR$(H \to \gamma\gamma) \approx 6 \times 10^{-6}$ \cite{susy-hit,CP-superH}. Thus, one has a diphoton production cross section~$\sigma (gg\to \gamma\gamma)$, when the resonant $s$-channel $H$- and $A$-boson exchanges are added, of about $\sigma (gg\to \Phi) \times {\rm BR}(\Phi \to \gamma\gamma) \approx 10^{-2} \; {\rm fb}$. Evidently, this cross-section value is at least two orders of magnitude too short of what was needed to explain the LHC diphoton excess, if this were due to the presence of new resonance(s).
  
The crucial question is therefore whether contributions of supersymmetric particles can generate such a huge enhancement factor of $\sim\! 100$. The chargino ($\chi^\pm_1$) contributions can be sufficiently large only in a rather contrived scenario, in which the mass~$m_{\chi_1^\pm}$ satisfies the relation $m_{\chi_1^\pm}\!=\! \frac12 M_A$ within less than a~MeV accuracy, such that a large factor of QED-corrected threshold effects can occur \cite{threshold}. In such a case, however,  finite-width regulating effects due to ($\chi^+_1\chi^-_1$) bound states might become important and may well invalidate this possibility. Here, we consider a more robust scenario, where the enhancement of the signal is driven mainly by a large $H\tilde{t}^*_1\tilde{t}_1$ coupling thanks to a large $\mu$ parameter, and the impact of possible  bound-state effects due to a~stoponium resonance~$\Sigma_{\tilde{t}}$ is properly assessed.

At leading order, the contributions of the top quark $t$ and its superpartners $\tilde t_1$ and $\tilde t_2$ to both the $H \gamma \gamma$ and $Hgg$ vertices\footnote{Because of CP invariance, the CP-odd boson $A$ does not couple to identical sfermions, so their quantum effects on  $Agg$ and $A\gamma\gamma$ vertices appear first at two loops and are therefore small. Note that the contributions of the first and second generation sfermions are  tiny while those of third generation sbottoms and staus are important only at very high $\tb$ values; they will all be included in the numerical analysis. The chargino loops in $\Phi \! \to \! \gamma\gamma$ can be neglected if there is no threshold enhancement \cite{threshold}.} (in our numerical analysis, all fermion and sfermion loops are included) are given by the amplitudes (up to  colour and electric charge factors)
\begin{eqnarray}
  \label{eq:phigg}
{\cal A} (H \gamma\gamma)\, \approx\, {\cal A} ( Hgg)\,\approx\, A_{1/2}^H (\tau_t) \times \cot\beta
+ \sum_{i=1,2} \hat g_{H \tilde t_i  \tilde t_i}/m_{\tilde t_i}^2  \times  A_{0}^H (\tau_{\tilde t_i})\,,
\end{eqnarray}
where the functional dependence of the form factors $A_{1/2}^H (\tau_i)$  and $A_{0}^H (\tau_i)$ for spin--$\frac12$ and  spin--0 particles (with $\tau_i=M_H^2/4m_i^2$ for the $i$th particle running in the loop) is displayed on the left pannel of Fig.~1.
As expected, they are real below the $M_H \!= \! 2m_i$ mass threshold and develop an imaginary part above this. The maxima are attained near the $t\bar{t}$ and $\tilde{t}^*_1\tilde{t}_1$-mass thresholds for the loop functions Re($A^H_{1/2})$ and Re($A^H_0)$, respectively. Specifically, for $\tau_i \!= \! 1$, one has ${\rm Re} (A^H_{1/2}) \approx  2.3$ and ${\rm Re}(A^H_{0})  \approx \frac43$, whilst Im($A^{H}_{0}) \approx 1$ for $\tau_i$ values slightly above the kinematical opening of the $\tilde{t}^*_1\tilde{t}_1$ threshold. Hence, the stop contribution is maximal for $m_{\tilde t_1} =  375$ GeV and, it is in fact comparable to the top quark one, since for $\tau_i \! = \!  M_H^2/4m_t^2 \! \approx \! 4.75$, one has $|A^H_{1/2} (\tau_t)| \! \approx \! 1.57$ to be contrasted with $|A^H_{0} (1)| \! \approx \! 1.33$. Since the SUSY scale $M_S \equiv \sqrt { m_{\tilde t_1} m_{\tilde t_2} }$ is supposed to be close to 1 TeV from naturalness arguments, one needs a large splitting between the two stops; the contribution of the heavier $\tilde t_2$ state to the loop amplitude is then small.  The significant stop-mass splitting is obtained by requiring a large mixing parameter which appears in the stop mass matrix, $X_t \! = \! A_t \! - \! \mu/\tan\beta$. At the same time, a large value of $X_t \!  \approx \! \sqrt 6 M_S$, together with $\tb \! \gsim \! 3$, maximize the radiative corrections to the mass $M_h$ of the observed $h$-boson\footnote{This scenario is reminiscent of the ``gluophobic" one discussed in Ref.~\cite{hstop} for the light $h$ boson but, here, the squark $\tilde t_1$ is rather heavy compared to $M_h$ and will have only a limited impact on the loop induced $gg \to h$ production and $h\to \gamma\gamma$ decay processes.}    and allows it to reach 125 GeV for a SUSY-breaking scale $M_S\sim 1$ TeV~\cite{benchmark,CPX4LHC}.

\begin{figure}[!h]
\vspace*{-1.6cm}
\centerline{ \hspace*{-1.5cm}
\includegraphics[scale=0.55]{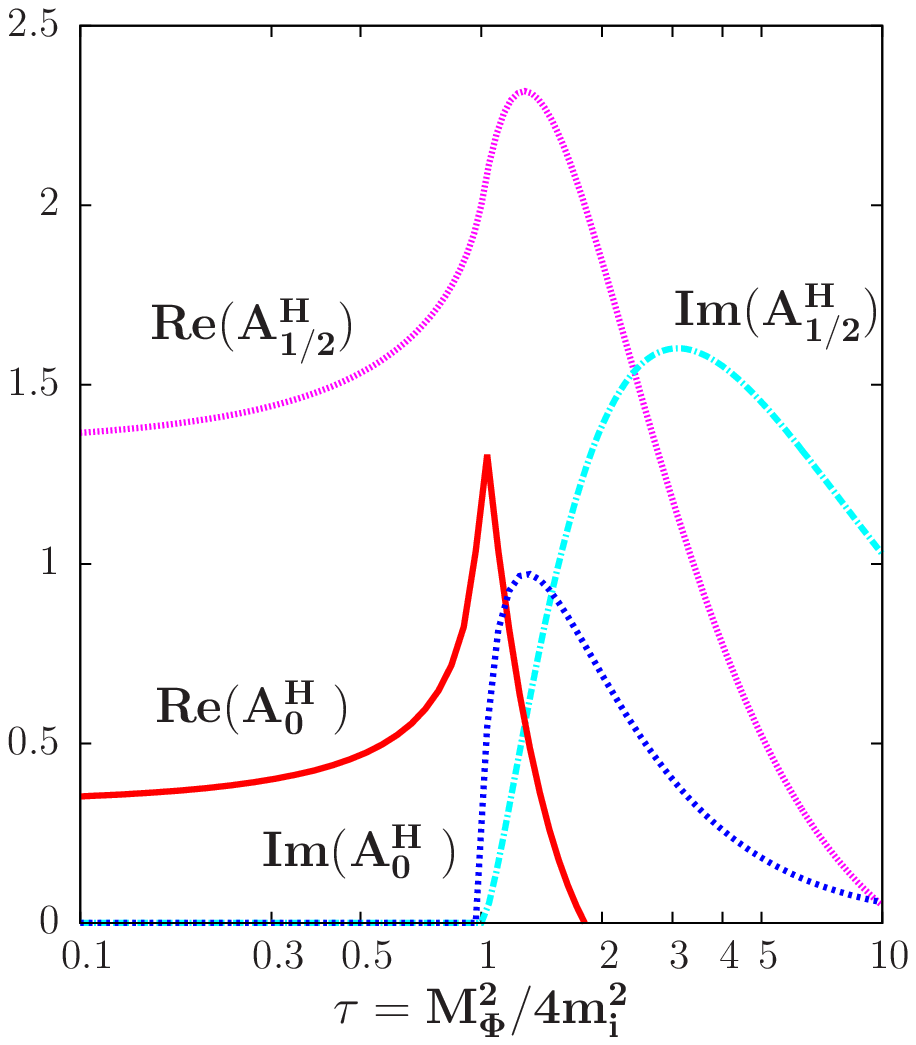} \hspace*{-6.2cm}
\includegraphics[scale=0.55]{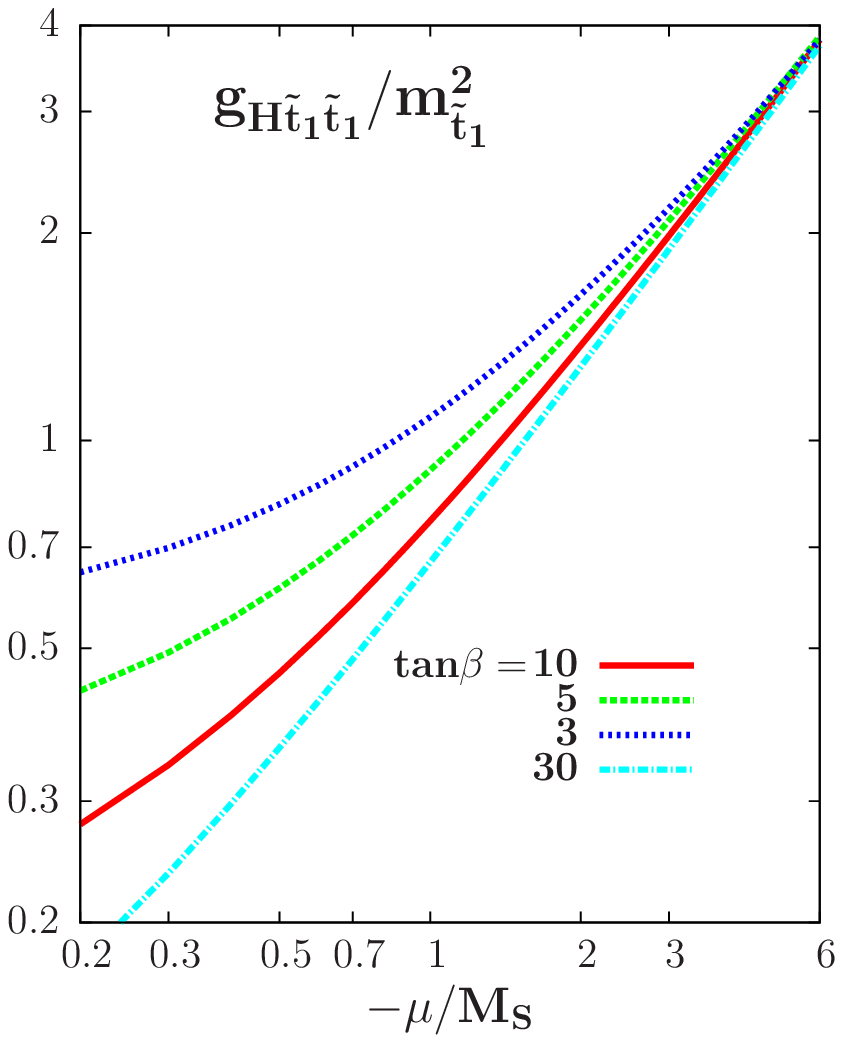} \hspace*{-6.5cm}
\includegraphics[scale=0.55]{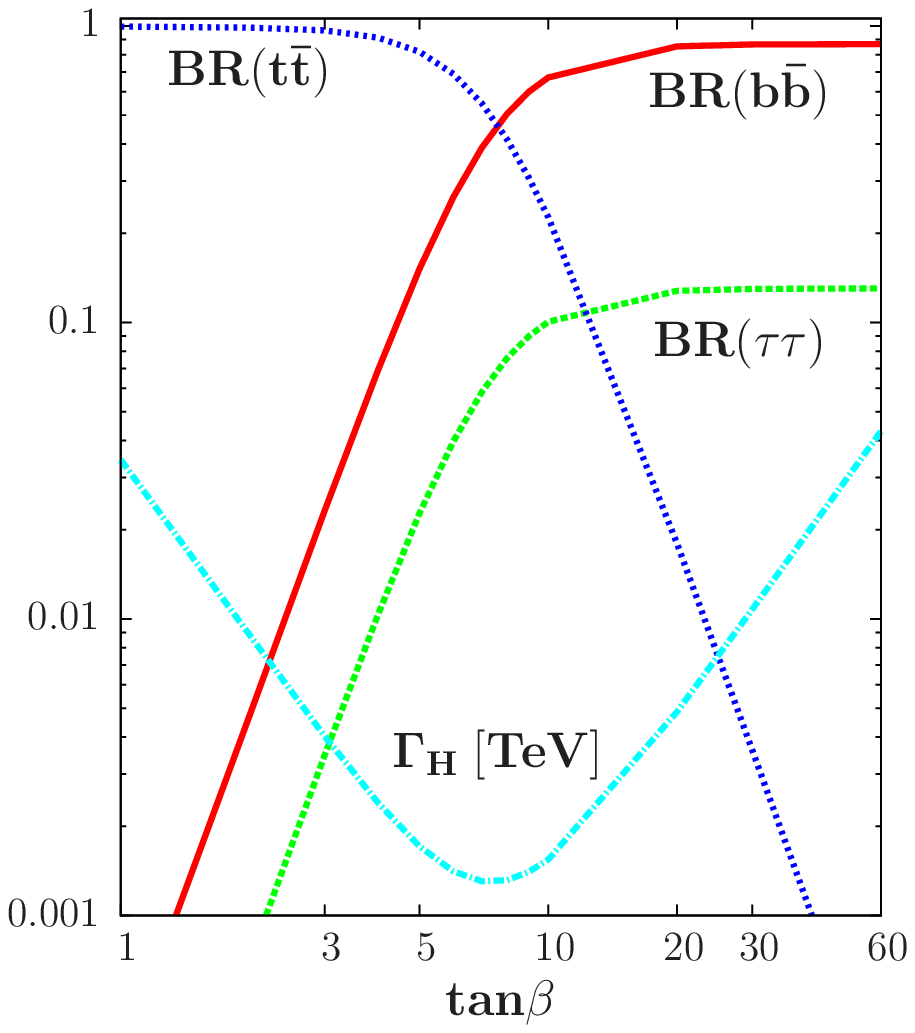}
}
\vspace*{-8.6cm}
\caption{\it Left: The real and imaginary parts of the form factors $A^H_{1/2}$ with fermion loops and $A^H_{0}$ with scalar loops as functions of the variable $\tau=M_H^2/4m_i$.
Center: the coupling (in units of $e M_Z/\cos\theta_W \sin\theta_W$) vs squared mass ratio $\hat{g}_{H \tilde t_1  \tilde t_1}/m_{\tilde t_1}^2$ as a function of the higssino mass $\mu$ [in TeV] for $A_t= \sqrt 6 M_S$ (with $M_S = 1$~TeV) and several $\tb$ values. Right: the $t\bar t, b\bar b, \tau \tau $ branching fractions and the total width [in TeV] of the $H$ state (when only decays into
fermions occurs) for $M_H=750$ GeV as a function of $\tb$.} 
\label{fig:enhancement}
\vspace*{-3mm}
\end{figure}

Large values of $\mu$ and $A_t$ (and of~$X_t$) increase considerably the $H \tilde t^*_1 \tilde t_1$ coupling that can strongly enhance the $Hgg$ and $H\gamma\gamma$ amplitudes. In the decoupling limit and for maximally mixed $\tilde t_i$ states, the tree-level $H \tilde t^*_1 \tilde t_1$-coupling is given by~\cite{Review} 
\begin{equation}
   \label{gstop}
\hat g_{H \tilde t_1 \tilde t_1} =\,  \frac{M_Z^2}{4}\sin2\beta + \cot^2\beta\; m_t^2 + \frac12 m_t 
(A_t \cot\beta -\mu)\,.
\end{equation}
In the central pannel of Fig.~1, $\hat g_{H \tilde t_1 \tilde t_1}$ is plotted as a function of $\mu$ for several values of $\tb$ and fixed $X_t = A_t - \mu \cot\beta = \sqrt 6\,M_S$, so as to get $M_h \approx 125$~GeV with a scale $M_S=1$~TeV. As can be seen from the central pannel, $\hat g_{H \tilde t_1 \tilde t_1}$ can be very large for $\mu$ in  the multi-TeV range. In fact, above the value $\tb \approx 3$, only the third term of~eq.~(\ref{gstop}) is important and the coupling is enhanced for large values of~$\mu$.  For instance, if $M_S \approx 1$~TeV and $m_{\tilde t_1}=375$~GeV, the $\tilde t_1$  contribution to the loop amplitudes in~eq.~(\ref{eq:phigg}) is roughly
\begin{equation} 
\hat g_{H \tilde t_1 \tilde t_1}/m_{\tilde t_1}^2 \times A_{0}^H (\tau_{\tilde t_1}) \approx - \frac{2}{3}\: \frac{m_t}{m_{\tilde t_1}}\: \frac{\mu}{m_{\tilde t_1}} \approx -0.8\, \frac{\mu}{M_S} \;.
\end{equation}
In particular, for $\mu = - 4M_S$ as in the so-called CPX scenario~\cite{CPX,CPX4LHC}, the stop effects can be twice as large as the top ones with $\tb =1$. This gives a prediction for the diphoton cross section which is about $2^4= 16$ times larger than that obtained for $\tb =1$.

Finally, one should take into account the size of the resonance width $\Gamma_H$. Indeed, the diphoton rate is given by the $gg\to H$ production cross section times the $H\to \gamma\gamma$ decay branching ratio and the impact of the total width~$\Gamma_H$ is important in the latter case. For $\tb\! =\!1$, the total width is almost exclusively generated by the $H \! \to \! t\bar t$ partial width, $\Gamma_H \approx \Gamma(H \! \to \! t\bar t) \! \propto \! m_t^2 \cot^2\beta /v $ and is about 30 GeV for $M_H\! =\! 750$ GeV. In our case, this situation is unacceptable since, as we have increased $\sigma( gg\to H)$ by including the stop contributions and we have BR$(H \! \to \! t \bar t) \! \approx \! 1$, $\sigma(gg \! \to \! H\! \to \! t\bar t)$ would be far too large and so is excluded by $t\bar t$ resonance searches \cite{LHC-tt}. BR($H\to t \bar t)$ needs thus to be suppressed and, at the same time,  also the total decay width which leads to an increase of BR$(H \! \to \! \gamma\gamma)$. This can be achieved by considering larger $\tb$ values for which
\begin{equation}
\Gamma_H  \propto \frac{ m_t^2}{v} \, \cot^2\beta +  \frac{ \bar m_b^2}{v}   \tan^2\beta + \frac{ m_\tau^2}{v}  \tan^2\beta \,.
\end{equation}
For $\tb \! =10$, one then obtains $\Gamma_H \! \approx \!2$ GeV and BR$(H \! \to \! \bar t t)\approx 20\%$ as can be seen in the right-hand side of Fig.~1, where the $H$ fermionic branching ratios and the total width are displayed as a function of $\tb$. The ratio BR$(H \to \gamma \gamma)$ can be thus increased, in principle,  by an order of magnitude compared to the $\tb\! =\! 1$ case.  Nevertheless, if a larger decay width is required for the resonance, one can increase the chosen $\tb$ value to, say $\tb \approx 20$ (i.e.~closer to the limit allowed by $H/A \to \tau \tau$ searches \cite{LHC-tautau}) and enhancing the $\tilde t_1$  contribution by increasing the value of the parameter $\mu$. However, values $\Gamma_H \! \gsim \! 30$ GeV cannot be achieved in principle\footnote{In fact, to obtain a sizeable total width, one option  could be to take $m_{\tilde t_1}$ a few GeV below the $\frac12 M_H$ threshold: one then opens  the $H\to \tilde t_1 \tilde t_1$ channel which increases the width $\Gamma_H$. This channel would suppress BR$(H \! \to \! \bar t t)$ as required at low $\tb$ but also BR$(H \! \to \! \gamma\gamma)$. Nevertheless, in the later case, some compensation can be obtained as the stop loop amplitude can be enhanced relative to the top one.}. Note that small values of $\tb$, $\tb \lsim 5$, cannot be tolerated, as they do not suppress enough BR$(H \! \to \! \bar t t)$ to a level to be compatible with $t\bar t$ resonance searches \cite{LHC-tt}.

\begin{figure}[!t]
\vspace*{-2.3cm}
\centerline{ \hspace*{-2cm} \includegraphics[scale=0.73]{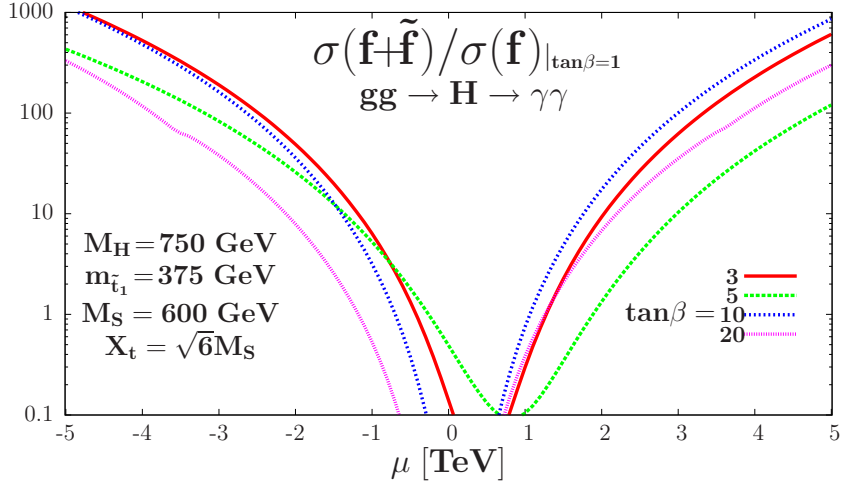}  }
\vspace*{-13.cm}
\caption{\it The enhancement factor of the diphoton cross section $\sigma(gg\to H) \times {\rm BR}(H \to \gamma \gamma)$ at the 13 TeV LHC as a function of $\mu$ [in TeV] for several values of $\tb$. It is obtained when including in the $Hgg$ and $H\gamma\gamma$ vertices third generation fermion~$f$ and all sfermion~$\tilde{f}$ loops,  in particular that of the lightest top squark $\tilde t_1$ with $m_{\tilde t_1}\! = \! \frac12 M_H \!\approx 375$~GeV, and is normalised to the rate when only the top quark loop is  present with $tan \beta=1$.}
\label{fig:plot-all}
\vspace*{-3mm}
\end{figure}

When all the ingredients discussed above are put together, the cross section $\sigma(gg\to H)$ times the decay branching ratio BR$(H \to \gamma \gamma)$ at the LHC with $\sqrt s=13$ TeV is displayed in Fig.~\ref{fig:plot-all} as a function of the parameter $\mu$ for the representative values $\tb=3,5,10, 20$ and $M_H\!=\!750$ GeV. The rate is normalised to the case where only top quark loops are present with $\tb=1$. The scenario features a light stop with $m_{\tilde t_1} \approx \frac12 M_H \approx 375$~GeV, which is obtained for a SUSY scale $M_S \approx 600$~GeV and a stop mixing parameter $X_t = \sqrt 6 M_S$, respectively. The contributions of the other states, the heavier stop with $m_{\tilde t_2} \simeq M_S^2/m_{\tilde t_1} \gsim 800$~GeV, the two sbottoms with $m_{\tilde b_{1,2}} \approx M_S$ (with couplings to the $H$ boson that are not enhanced) and the first/second generations sfermions (assumed to be much heavier than 1 TeV) are included together with the ones of the bottom quark, but they are small compared to that of the lightest $\tilde t_1$. As can be seen, for $\tb=10$ for instance, an enhancement by a factor of about 100 can be obtained for a value $|\mu | =3$ TeV, i.e. $|\mu | \simeq 5 M_S$~\footnote{Such large values of $|\mu |$ can be obtained, for instance, in the context of the new MSSM~\cite{nMSSM}, in which the tadpole term $t_S S$ for the singlet field $S$ has different origin from the soft SUSY-breaking mass~$m^2_S S^*S$. For values of $t^{1/3}_S \gg m_S$, a large vacuum expectation value for $S$ can be generated of order $v_S \equiv \langle S\rangle \simeq t_S/m^2_S \gg m_S \sim M_S$, giving rise to a large effective $\mu$ parameter: $\mu_{\rm eff} = \lambda v_S \gg M_S$, where $\lambda \lsim 0.6$ is the superpotential coupling of the chiral singlet superfield $S$ to the Higgs doublet superfields $H_u$ and $H_d$.
  Hence, in this new MSSM setting, the appearance of potentially dangerous charged- and colour-breaking minima~\cite{CCB} due to a large $\mu_{\rm eff}$ can be avoided more naturally than in the MSSM.}. Hence, one easily arrives at production cross sections of ${\cal O}(1~{\rm fb})$ at the LHC for heavy Higgs resonances well above the $t\bar{t}$ threshold decaying sizeably into two photons, e.g.~comparable to the diphoton cross sections initially observed by ATLAS and CMS~in their early 13 TeV data \cite{diphoton}.

While the $M_S$ and $X_t$ values adopted for the figure above lead to sufficiently large radiative corrections to generate a mass for the lighter $h$ state that is close to $M_h=125$~GeV for $\tb \gsim 5$ (in particular if an uncertainty of a few GeV from its determination is taken into account \cite{benchmark}), the required large $\mu$ value might be problematic in some cases. Indeed, at high $\tb$ and $\mu$,  there are additional one--loop vertex corrections that modify the Higgs couplings to $b$--quarks, the dominant components being given by \cite{CR-deltab}
\begin{equation}
   \label{eq:Db}
\Delta_b \simeq  \bigg(\frac{2\alpha_s}{3\pi}\; \frac{m_{\tilde{g}} }{{\rm max}(m_{\tilde{g}}^2, m_{\tilde{b}_1}^2, m_{\tilde{b}_2}^2)} + \frac{\lambda_t^2}{16\pi^2}\; \frac{ A_t}{{\rm max}(\mu^2, m_{\tilde{t}_1}^2, m_{\tilde{t}_2}^2)} \bigg)\, \mu \tb\, ,
\end{equation}
where $\lambda_t = \sqrt 2 m_t/v$. Note that in eq.~(\ref{eq:Db}),  the first and second terms are the dominant gluino--sbottom and stop--chargino loop corrections to the $Hb\bar b$ coupling, respectively. For~$|\mu| \tb \gg M_S$, as is required here, these corrections become very large and would, for instance, lead to an unacceptable value for the bottom quark mass. Hence, either 
one should keep $|\mu| \lsim 5M_S$ or alternatively, partly or fully cancel the two terms of the equation above. This, for instance, can be achieved by choosing a trilinear coupling $A_t <0$ and a very heavy gluino with a mass $m_{\tilde{g}}$ such that $m_{\tilde{g}} \approx - 4|\mu|^2/ A_t$. 

Nevertheless, the leading order discussion held above is not sufficient to address all the issues involved in this context and it would be desirable to provide accurate predictions for a ``realistic" MSSM scenario, for which all important higher order effects are consistently implemented as in one of the established public codes. Specifically, using the program {\tt SUSY-HIT} \cite{susy-hit} which calculates the spectrum (through {\tt Suspect}) and decays (through {\tt HDECAY} and {\tt SDECAY}) of the Higgs and SUSY particles, we have identified MSSM benchmark points in which the $gg\to H\to \gamma\gamma$ rate is almost entirely explained when NLO QCD corrections to the rate are included as in Ref.~\cite{stop-qcd}. For instance, for $\tb\!=\!10$, third--generation scalar masses of $m_{\tilde t_L}\!=\! m_{\tilde t_R} \! = \! m_{\tilde b_R} \! = \! 0.8$~TeV $\approx \! M_S$, trilinear couplings $A_t \! = \! A_b \! = \! 2$~TeV, gaugino mass parameters $M_1 \! = \! \frac12 M_2 \! = \! \frac16 M_3 \! = \! 350$~GeV and a higgsino mass $\mu \! = \! 2.3$~TeV, the program {\tt Suspect2} (version 2.41) yields $m_{\tilde t_1}= 373.75$ GeV and $m_{\tilde t_2}=847$~GeV. Moreover, fixing the CP--odd $A$-scalar mass to $M_A=756$~GeV, one obtains $M_H=747.6$~GeV, which is somewhat above the $2m_{\tilde t_1}\! \approx\! 747.5$~GeV threshold, and $M_h=121$~GeV, but with an inherent theoretical uncertainty estimated to be~$\sim\! 3$--4~GeV. With these inputs, an enhancement factor of at least two orders of magnitude is obtained, when compared to the case in which only the top loop contributes with $\tb=1$. In detail, {\tt HDECAY 3.4} computes BR($H \! \to \! \gamma\gamma) \!= \! 9.2 \! \times \! 10^{-4}$, BR($H \! \to \! gg) \!= \! 4.2 \! \times \! 10^{-2}$ and a total width $\Gamma_{H} \! = \! 2.06$~GeV, to be compared with BR($H \! \to \! \gamma\gamma) \! = \! 6 \! \times \! 10^{-6}$, BR($H \! \to \! gg) \! = \! 1.8 \! \times \! 10^{-3}$ and $\Gamma_{H} \! = \! 35$~GeV without stop loops and $\tb\!=\!1$. Hence, making the plausible assumption that the QCD corrections vary the same way in both the $gg$ production and decay rates, we get an enhancement factor of $\sim 200$, leading to a cross section $\sigma(gg\!\to \! H\! \to \! \gamma\gamma) \approx 0.83$~fb at the LHC with $\sqrt s=13$ TeV. Also, we expect additional contributions to come from other sources, as we will discuss below.  Note that besides giving rise to an $h$~boson with a mass $M_h$ close to 125~GeV and SM-like couplings, this benchmark point leads to BR($H\to \tau\tau)=7\%$ and BR$(H\to t\bar t)=15\%$.  Given that only the $gg \to H$ production channel gets enhanced thanks to~$\tilde{t}_1\tilde{t}^*_1$-threshold effects, we can thus estimate that $\sigma(gg\to H)\; \mbox{BR}(H\to \tau\tau ) \approx 62$~fb at $\sqrt{s}=13$~TeV, which satisfies the current LHC limits deduced from direct MSSM Higgs searches in the $\tau\tau$ final state~\cite{Aaboud:2016cre}, in particular when one takes into account the uncertainty bands reported there.  The complete input and output program files for the aforementioned benchmark point are available upon request\footnote{We thank Pietro Slavich for his cooperation on this issue.}.

Two additional sources of corrections might significantly increase the $gg\!\to \! H\! \to \! \gamma\gamma$ production cross section, as we will briefly outline below, and need to be taken into account.

The first one is that the form factor for the $H\gamma\gamma$ and $Hgg$ couplings appearing in eq.~(\ref{eq:phigg}) and displayed in Fig.~\ref{fig:enhancement} (left) does not accurately describe the threshold region, $m_{\tilde{t}_1}\approx \frac12 M_H$~\cite{stop-qcd} that we are interested in here\footnote{Our estimates are performed by defining all input parameters in the $\overline{\rm DR}$ scheme, including the stop masses~$m_{\tilde{t}_{1,2}}$ and the trilinear Yukawa coupling~$A_t$. To accurately address, however, the issue of threshold and stoponium effects, other IR-safe renormalization schemes may be more appropriate, especially for the definition of the coloured
$\tilde{t}_1$-particle mass $m_{\tilde{t}_1}$, similar to the potential-subtracted and 1S renormalization schemes for the $t$-quark mass~$m_t$ used in higher-order computations of $t\bar{t}$~production at threshold~\cite{mtopdef}. However, such scheme redefinitions for $m_{\tilde t_1}$ and $A_t$ do not generally change the predicted values of physical observables, such as decay rates and cross sections, at a given loop order of the perturbation expansion.}. This is because  when the stop mass lies slightly above threshold, a Coulomb singularity develops signalling the formation of S--wave (quasi) bound states~\cite{Bigi,Coulomb,MSY}. Following Ref.~\cite{threshold}, this can be taken into account, in a non-relativistic approach\footnote{In the context of QCD, we are dealing with a region in the deep infra-red regime where non--perturbative gluon mass effects that extend up to the GeV region might be needed to be taken into account \cite{papa}. In view of the lack of first principle's calculation for the case of quasi--stable top squarks, we perform a conservative  estimate by adapting the results of the non--relativistic approach in~\cite{MSY}.},  by re-writing the form factor close to threshold as~\cite{MSY}
\begin{equation}
A_{0}^H =a+b \times G(0,0;E_{\tilde{t}_1} +i \Gamma_{\tilde{t}_1}^{\rm eff}),
\end{equation}
where $a$ and $b$ are perturbative calculable coefficients obtained from matching the non--relativistic theory to the full theory. To leading order, one has $a=\frac12 (1-\frac{\pi^2}{4})$ and $b= 2 \pi^2/m_{\tilde{t}_1}^2$ for the real and imaginary parts, respectively. Moreover, $E_{\tilde{t}_1} \! =\! M_H \! -\! 2 m_{\tilde t_1}$ is the energy gap from the threshold region and $\Gamma_{\tilde{t}_1}^{\rm eff}$ is a regulating effective scattering width for the top squark in the loop which can be of ${\cal O}$(1 GeV) or below. If the stop total width~$\Gamma_{\tilde{t}}$ happens to be too small, specifically if~$\Gamma_{\tilde{t}}\ll 1$~GeV, $\Gamma_{\tilde{t}_1}^{\rm eff}$ is expected to be then of order the decay width~$\Gamma_{\Sigma_{\tilde{t}}}$ of the stoponium state $\Sigma_{\tilde{t}}$ whose impact on the diphoton excess will be discussed later. Finally, $G(0,0;E_{\tilde f})$ is the S--wave Green's function of the non-relativistic Schr\"{o}dinger equation in the presence of a Coulomb potential $V(r)=-C_F\alpha/r$~\cite{GreensFn}.

Following Ref.~\cite{threshold}, we have estimated the absolute value of the enhancement factor $F$, defined as $F={A^H_{0}(\mbox{threshold enhanced})}/{A^H_{0}(\mbox{perturbative})}$, as a function of the effective width $\Gamma_{\tilde{t}_1}^{\rm eff}$, for a resonance mass $M_H=750$ GeV and an energy gap $E_{\tilde{t}_1} \!= \! M_H \! - \! 2 m_{\tilde{t}_1}$ negative and of order~1 GeV. We find\footnote{We thank Aoife Bharucha for her help in this issue.} that for $\Gamma_{\tilde{t}_1}^{\rm eff} = \Gamma_{\tilde{t}_1} = {\cal O}(1\;$GeV), one can easily obtain an enhancement factor of~2, while for a smaller $\tilde t_1$ decay width, a much larger factor is possible. For instance, for $\Gamma_{\tilde{t}_1} \approx 200$ MeV (which can easily be achieved if the mass difference between $\tilde t_1$ and the lightest neutralino $\chi_1^0$ is small enough so that only three-- or four--body or loop induced $\tilde t_1$ decay modes can occur), the enhancement factor in the $H \! \to \! \gamma\gamma$ amplitude is about $\!2,\, 4,\, 8$, for $E_{\tilde{t}_1}\!= \!-1.5, \!-\! 2, \! -\! 2.5$ GeV, respectively. Note that the maximum enhancement of a factor~8 is reached for $E_{\tilde{t}_1} \approx -2.5$~GeV.
 
Hence, considering that a similar threshold enhancement could be present in the $Hgg$ amplitude, one can achieve at least one order of magnitude enhancement in the $gg\to H \to \gamma\gamma$ cross section times branching ratio compared to the leading order result. Together with the initial one loop contribution of the $m_{\tilde t_1} \approx \frac12 M_H$ top squark discussed before, this will be sufficient to increase the diphoton production rate to the level of ${\cal O}(1~{\rm fb})$.  In addition, possible QCD threshold enhancements can be sufficiently large so as to avoid considering too high $\mu$ or $A_t$ values to enhance the coupling $g_{H \tilde t_1 \tilde t_1}$, and one can thus obtain sizeable diphoton production cross sections of ${\cal O}(1~{\rm fb})$ at the LHC, even with basic SUSY parameters that can occur in constrained MSSM scenarios, such as the minimal supergravity model with non--universal Higgs mass parameters~\cite{MSSM}.

A second important issue that needs to be addressed is the formation of the stoponium bound states $\Sigma_{\tilde{t}}$ and their mixing with the CP-even $H$ boson\footnote{As this work was being finalized for submission, Ref.~\cite{stoponia} appeared in which the stoponium bound state was put forward as the only source for an enhanced diphoton rate of the size reported in~\cite{diphoton}. There is some partial overlap with our discussion here but the mixing with the $H$ boson, and more generally all issues related to this Higgs state (which is almost entirely responsible of the diphoton excess in our case), have not been considered in Ref.~\cite{stoponia}.}. For our illustrations, we only consider the lowest lying 1S scalar stoponium state~$\Sigma_{\tilde t}$, which can mix with the $H$ boson. Our approach is similar to Ref.~\cite{Stoponium}, and we ignore the potential impact of $s$-dependent effects on the $H$ and $\Sigma_{\tilde t}$ masses, their widths and their mixings~\cite{APNPB}. In this simplified scenario, the resonant transition amplitude ${\cal A}_{\rm res}(s) = {\cal A}(gg \to H, \Sigma_{\tilde t} \to \gamma\gamma )$ is given by
\begin{equation}
{\cal A}_{\rm res}(s)\ =\ \left( {\cal V}^g_H\,,\ 
{\cal V}^g_{\Sigma_{\tilde{t}}} \right)\, 
\left(\!\begin{array}{cc}
s - M^2_H + i M_H \Gamma_H & \delta M^2_{H\Sigma_{\tilde{t}}} \\
\delta M^2_{H\Sigma_{\tilde{t}}}  & s - M^2_{\Sigma_{\tilde{t}}} + i M_{\Sigma_{\tilde{t}}} \Gamma_{\Sigma_{\tilde{t}}}\\
\end{array}\!\right)^{-1}\,
\left(\!
\begin{array}{c}
{\cal V}^\gamma_H\\
{\cal V}^\gamma_{\Sigma_{\tilde{t}}} 
\end{array}\!\right)\; ,
\end{equation}
where ${\cal V}^g_H$ (${\cal V}^\gamma_H$) and ${\cal V}^g_{\Sigma_{\tilde{t}}}$ (${\cal V}^\gamma_{\Sigma_{\tilde{t}}}$) are the effective couplings of $H$ and $\Sigma_{\tilde{t}}$ to the gluons~$g$ (photons~$\gamma$), and we neglect non-resonant contributions in our estimates.  For the lowest lying state~$\Sigma_{\tilde{t}}$, its mixing $\delta M^2_{H\Sigma_{\tilde{t}}} $ with the $H$ boson is purely dispersive and of ${\cal O}(40~{\rm GeV})\times M_{\Sigma_{\tilde{t}}}$, as estimated in the Coulomb approximation, by virtue of  eqs.~(3.10)--(3.12) of~\cite{Hagiwara}. Moreover, we observe that for $\tan\beta\!\sim\!5$--10, the decay widths of the heavy $H$ boson and the stoponium~$\Sigma_{\tilde{t}}$ are comparable in size, i.e., $\Gamma_H \sim \Gamma_{\Sigma_{\tilde{t}}} \sim {\cal O}({\rm GeV})$~\cite{Stoponium}, but the effective $H$ couplings ${\cal V}^{g,\gamma}_H$ are QCD-enhanced with respect to the $\Sigma_{\tilde{t}}$ couplings ${\cal V}^{g,\gamma}_{\Sigma_{\tilde{t}}}$ by a factor of~2 (or more). Consequently, the amplitude ${\cal A}_{\rm res}(s)$, with only the $H$-boson included, is at least a factor of 3 larger than the one with only the stoponium~$\Sigma_{\tilde{t}}$ being  considered.

At the cross section level, we may naively estimate that ignoring potentially destructive Higgs-stoponium interference effects~\cite{Bodwin:2016whr}, the inclusion of all stoponium resonances can increase the signal cross section~$\sigma (gg \! \to \! \Phi \! \to \! \gamma\gamma)$ by up to a factor of 1.5, especially if one adopts the results for the stoponium wave-function $R_{nS}(0)$ at the origin, from non-relativistic lattice computations~\cite{Kim:2015zqa}. This increase in the signal rate would open a somewhat wider portion of the MSSM parameter space for an enhanced production rate of diphoton resonances.

In summary, in this exploratory Letter we have considered scenarios in the context of the MSSM
in which very large diphoton rates can be obtained at the current and future LHC runs. For the sake of illustration, we have taken the example of the excess in the diphoton spectrum observed by ATLAS and CMS in their early 13 TeV data \cite{diphoton} and which turned out to be simply a statistical fluctuation \cite{LHC2}. In the context of the MSSM, this excess of ${\cal O}$(fb)
could have been explained by the production of the heavier CP--even $H$ boson of a mass $M_H \! \simeq \! 750$ GeV, with the large $gg \! \to \! H$ production cross section times $H \! \to \! \gamma \gamma$ decay branching ratio. This enhancement is a combination of three different sources, all related to the fact that the lighter top squark\footnote{A similar mechanism with light bottom squarks can be invoked but it is disfavoured compared to the stop one because: (i)~the electric charge $e_b=-\frac13$ forces us to pay a penalty of a factor 4 in the $H\gamma\gamma$ vertex and (ii)~it is more difficult to enhance the $H \tilde b^*_1 \tilde b_1$ coupling to the required level, since $\hat{g}_{{H \tilde b_1 \tilde b_1}}\! \propto \! m_b (A_b \tan\beta \! - \! \mu)$. For the case of $\tau$-sleptons, the situation is even worse as they affect only the $H \gamma \gamma$ loop and the relevant coupling $\hat{g}_{{H \tilde \tau_1 \tilde \tau_1}}$ is smaller by a factor $m_b/m_\tau$.} has a mass close to the $\frac12 M_H$ threshold, i.e. $m_{\tilde t_1}\approx 375$ GeV.  The first one is that, at leading order, $\tilde t_1$ contributes maximally to the $Hgg$ and $H\gamma\gamma$ amplitudes, especially if the $H\tilde t^*_1 \tilde t_1$ coupling is strong which can be achieved by allowing large values for the higgsino mass parameter $\mu$. Compared to the case where only the top quark contribution is considered for $\tan\beta\!=\!1$ for which it is maximal, an enhancement factor of two orders of magnitude for the $gg\! \to \! \Phi \! \to \! \gamma\gamma$ signal can be achieved. This alone, might be sufficient to obtain ${\cal O}$(fb)
diphoton rates.  Nevertheless, a second source of enhancement can come from the inclusion of QCD corrections to the $H \! \to \! \gamma\gamma$ process near the $\frac12 M_H$ threshold which can easily lead to an extra factor of~2 or more enhancement at the amplitude level. 
Finally, a last ingredient is the formation of stoponium bound states which can mix with the $H$ boson. Their effect might increase the $gg \! \to \! \Phi \! \to \! \gamma\gamma$ rate by another factor of about~2. Hence, the addition of these many enhancement factors will give rise to an enhanced diphoton cross section of ${\cal O}(1~{\rm fb})$ for a heavy diphoton Higgs resonance, having a mass well above the $t\bar{t}$ threshold, e.g.~with $M_H \approx 750$~GeV, even within the context of the plain MSSM\footnote{To the best of our knowledge, the present Letter and the earlier attempt in Ref.~\cite{threshold} have offered the first interpretations for 750 GeV diphoton resonances with enhanced production rates within the context of the usual MSSM with $R$-parity conservation and without any additional particle content.  Otherwise, other minimal beyond-the-MSSM suggestions include the $R$-parity violating MSSM~\cite{Allanach:2015ixl} and the next-to-MSSM~\cite{NMSSM}.}.

The scenario thus features light top and bottom squarks and, hence, a relatively low SUSY scale $M_S \lsim 1$ TeV as favoured by naturalness arguments. This nevertheless allows for the $h$-boson mass to be close to 125 GeV, if $\tb$ is relatively large and stop mixing maximal as in our case. In order to cope with constraints from SUSY particle searches at the LHC \cite{PDG}, the gluino and the first/second generation squarks  should have masses above the TeV scale. The charginos and neutralinos should also be heavy (in particular the higgsinos as $\mu$~is large) except  the lightest neutralino $\chi_1^0$, which could be the lightest stable SUSY-particle (LSP) and must have a mass only slightly lower than $m_{\tilde t_1}$, as LHC limits on $m_{\tilde t_1}$ are practically non-existent if $m_{\chi_1^0}\!  \gsim 300$~GeV~\cite{PDG}. In this case,  the first accessible visible SUSY state at the LHC would be $\tilde t_1$ which will mainly decay into $\tilde t_1 \to c  \chi_1^0$ (via loops) and  $\tilde t_1 \to b f \bar f'  \chi_1^0$ (at the three- or four-body level) \cite{stop1}. The dominant decays of the heavier stop\footnote{The rate for $pp\to \tilde t_2^* \tilde t_2 \to ZZ \chi_1^0 \chi_1^0 jj \to \ell^+ \ell^-$+jets+missing energy  would be in the right ballpark for $m_{\tilde t_2 }\approx 600$--800 GeV, so as to explain the apparent $3\sigma$ excess in the ATLAS data at $\sqrt s\! = \! 8$ TeV \cite{Zll-ATLAS}.}  will be $\tilde t_2 \to \tilde t_1 Z$ and to a lesser extent $\tilde t_2 \to \tilde t_1 h$, while those of two bottom squarks could almost exclusively be $\tilde b_{1,2} \to \tilde t_1 W$.  Hence, besides $M_H \! \approx \! 2m_{\tilde t_1}$ which is a firm prediction, the present scenario favours a light third generation squark spectrum, as well as the usual MSSM  degenerate heavy Higgs spectrum, $M_A \approx M_{H^\pm} \approx M_H$, that can be probed at the current LHC~run.  

Our scenario exhibits a number of other interesting phenomenological features that need to be discussed in more detail. On the Higgs side, for instance, one would like to precisely determine the impact of the SUSY particle spectrum on the tree--level and loop-induced decays of the MSSM Higgs states, such as $H \to Z\gamma$ in which similar effects as in $H \to \gamma\gamma$ might occur, as well as quasi--on--shell $H^{(*)} \to \tilde t_1^* \tilde t_1$ which offers a direct and falsifiable test of the actual threshold enhancement mechanism under study here. Another interesting issue would be to explore the possibility of resonant CP--violating effects at the $\Phi$ resonance which could then be a mixture of the CP--even and CP--odd states~\cite{APNPB}. In the case of the supersymmetric spectrum, our scenario leads to relatively light top and bottom squarks as discussed above and it would be interesting to study how they can be detected in the presence of, not only a bino--like LSP that is nearly mass degenerate with the $\tilde t_1$ state, but also a gravitino LSP in both gravity or gauge-mediated SUSY--breaking scenarios. This last aspect can have two important consequences: (i)~the $\tilde t_1$ total width would be very small, as only multi-body or loop-generated decays will be allowed \cite{stop1}, and (ii)~the relic density of the bino dark matter might be obtained through stop--neutralino co--annihilation \cite{stop2}.  

Hence, within the context of the MSSM, diphoton resonances produced with largely enhanced rates at the LHC could lead to an extremely interesting phenomenology both in the Higgs and the superparticle sectors. Some of these aspects have been briefly touched upon in this note and we leave the discussion of many other aspects to a forthcoming study~\cite{progress}. \bigskip

\noindent {\bf Acknowledgements:}  
We would like to thank Aoife Bharucha for collaboration at the early stages of this work.  
Discussions with Manuel Drees, Michael Spira and Pietro Slavich are gratefully acknowledged. AD is supported by the ERC advanced grant Higgs@LHC and AP by the Lancaster--Manchester--Sheffield Consortium for Fundamental Physics, under STFC research grant: ST/L000520/1.



\begin{thebibliography}{999}
\begin{small}

\bibitem{diphoton} ATLAS and CMS collaborations: ATLAS-CONF-2015-081; CMS-PAS EXO-15-004.\vspace*{-1mm}

\bibitem{LHC2} B. Lenzi (ATLAS) and M. Rovelli (CMS) talks at ICHEP in Chicago on 5 August 2016.\vspace*{-1mm}

\bibitem{phi} A complete list of papers dealing with the 750 GeV resonance can be obtained from: {\tt http://inspirehep.net/search?ln=en\&p=refersto\%3Arecid\%3A1410174}.\vspace*{-1mm}

\bibitem{MSSM} M. Drees, R. Godbole and P. Roy,  {\it Theory and phenomenology of sparticles}, World Scientific, 2005; H. Baer and X. Tata, {\it Weak scale Supersymmetry: from superfields to scattering events}, Cambridge U. Press, 2006; S. Martin, hep-ph/9709356.\vspace*{-1mm}  
 
\bibitem{Review} 
\mbox{J. Gunion, H. Haber, G. Kane and S. Dawson, ``The Higgs Hunter's
Guide", Reading 1990;} \\ 
A.~Djouadi, Phys. Rept. 459 (2008) 1.\vspace*{-1mm}

\bibitem{ADM} A. Angelescu, A. Djouadi and G. Moreau, Phys.~Lett.~B756 (2016) 126;\\
A. Djouadi, J. Ellis, R. Godbole and J. Quevillon,   JHEP 1603 (2016) 205.\vspace*{-1mm} 

\bibitem{PDG} Particle Data Group (K. Olive et al.), Chin. Phys. C38 (2014) 090001.\vspace*{-1mm} 
  
\bibitem{hMSSM} A. Djouadi et al., JHEP 06 (2015) 168;  Eur.~Phys.~J. C73 (2013) 2650;\\
A. Djouadi and J. Quevillon, JHEP 10 (2013) 028.\vspace*{-1mm}

\bibitem{Stoponium} M.~Drees and K.~Hikasa,
  Phys.\ Rev.\ D{41} (1990) 1547;\\
M.~Drees and M.~M.~Nojiri,
  Phys.\ Rev.\ D{49} (1994) 4595.\vspace*{-1mm}

\bibitem{LHC-tautau} ATLAS collaboration, JHEP 11 (2014) 056;  CMS collaboration, JHEP 10 (2014) 160.\vspace*{-1mm} 

\bibitem{LHC-tt} ATLAS collaboration, JHEP 08 (2015) 148; CMS collaboration, arXiv:1506.03062.\vspace*{-1mm}   

\bibitem{LHC-prod} S. Dittmaier et al., LHC Higgs cross section Working Group, arXiv:1101.0593; J. Baglio and A. Djouadi, JHEP 03 (2011) 055; M. Spira et al., Nucl. Phys. B453 (1995) 17.\vspace*{-1mm}

\bibitem{susy-hit}  A. Djouadi, J-L. Kneur and G. Moultaka, Comput. Phys. Commun. 176 (2007) 426;\\ A.~Djouadi, J.~Kalinowski and M.~Spira, Comput. Phys. Commun. 108 (1998) 56;\\ 
M. Muhlleitner, A. Djouadi and Y. Mambrini, Comput. Phys. Commun. 168 (2005) 46;\\
A. Djouadi, M. Muhlleitner and M. Spira, Acta. Phys. Polon. B38 (2007) 635.\vspace*{-1mm} 

\bibitem{CP-superH} J.~S.~Lee et al., Comput.\ Phys.\ Commun.\  156 (2004) 283; Comput.\ Phys.\ Commun.\  180 (2009) 312;  Comput.\ Phys.\ Commun.\ 184 (2013) 1220.\vspace*{-1mm} 

\bibitem{threshold} A. Bharucha, A. Djouadi and A. Goudelis, arXiv:1603.04464.\vspace*{-1mm}  

\bibitem{hstop} A. Djouadi,  Phys.~Lett.~B435 (1998) 101.\vspace*{-1mm} 

\bibitem{benchmark} M. Carena et al., Eur. Phys. J. C73 (2013) 2552.\vspace*{-1mm} 

\bibitem{CPX4LHC}  M.~Carena, J.R.~Ellis, J.S.~Lee, A.~Pilaftsis and C.E.M.~Wagner,
JHEP 1602 (2016) 123.\vspace*{-1mm} 

\bibitem{CPX} M.~Carena, J.R.~Ellis, A.~Pilaftsis and C.E.M.~Wagner, Phys. Lett. B495 (2000) 155.\vspace*{-1mm}

\bibitem{nMSSM}
  C.~Panagiotakopoulos and K.~Tamvakis,
  Phys.\ Lett.\ B469 (1999) 145;\\
C.~Panagiotakopoulos and A.~Pilaftsis,
  Phys.\ Rev.\ D63 (2001) 055003.\vspace*{-1mm} 

\bibitem{CCB} 
For example, see, J. Casas, A. Lleyda and C. Munoz, Nucl.~Phys.\ B471 (1996) 3;\\ 
J. Camargo-Molina, B. O'Leary, W. Porod and F. Staub, JHEP 12 (2013) 103.\vspace*{-1mm}

\bibitem{CR-deltab} See e.g., M.~Carena, D.~Garcia, U.~Nierste and C.~Wagner, 
Nucl.~Phys.~B577~(2000)~88;\\
 D. Noth and M. Spira, Phys. Rev. Lett. 101 (2008) 181801.\vspace*{-1mm}

\bibitem{stop-qcd} See for instance, M. Muhlleitner and M. Spira, Nucl. Phys. B790 (2008) 1.\vspace*{-1mm}  

\bibitem{Aaboud:2016cre} 
ATLAS Collaboration,
  Eur.\ Phys.\ J.\ C76 (2016) 585
  [arXiv:1608.00890 [hep-ex]].

\bibitem{mtopdef} M. Beneke,  Phys.~Lett.\ B434 (1998) 115;\\
A. H. Hoang and T. Teubner, Phys.~Rev.\ D60 (1999) 114027.\vspace*{-1mm}

\bibitem{Bigi}  I.~Bigi, V.~Fadin and V.~Khoze, Nucl. Phys. B377 (1992) 461.\vspace*{-1mm} 

\bibitem{MSY} K. Melnikov, M. Spira and O. Yakovlev, Z. Phys. C64 (1994) 401.\vspace*{-1mm}   

\bibitem{Coulomb} For a recent discussion, see for instance, M. Beneke et al., JHEP 03 (2016) 119.\vspace*{-1mm} 

\bibitem{papa} A.~C. Aguilar, D. Binosi and J. Papavassiliou, Phys. Rev. D78 (2008) 025010;\\
A.~C.~Aguilar, D.~Binosi, C.~T.~Figueiredo and J.~Papavassiliou, arXiv:1604.08456.\vspace*{-1mm} 

\bibitem{GreensFn} V.~Fadin and V.~Khoze, Sov.\ J.\ Nucl.\ Phys.\  {48} (1988) 309 
[Yad.\ Fiz.\  {48} (1988) 487]; JETP Lett.\  {46} (1987) 525~[Pisma Zh.\ Eksp.\ Teor.\ Fiz.\  {46} (1987) 417].\vspace*{-1mm} 

\bibitem{stoponia} D. Choudhury and K. Ghosh, arXiv:1605.00013.\vspace*{-1mm}  

\bibitem{APNPB} A.~Pilaftsis,
  Nucl.\ Phys.\ B{504} (1997) 61.\vspace*{-1mm}

\bibitem{Hagiwara} K. Hagiwara, K. Kato, A.~D. Martin and C.-K. Ng., Nucl.\ Phys.\ 344 (1980) 
1.\vspace*{-1mm} 

\bibitem{Bodwin:2016whr}
  G.~T.~Bodwin, H.~S.~Chung and C.~E.~M.~Wagner,
  arXiv:1609.04831 [hep-ph].

\bibitem{Kim:2015zqa}
  S.~Kim,
  Phys.\ Rev.\ D{92} (2015)  094505. \vspace*{-1mm}

\bibitem{Allanach:2015ixl}
  B.~C.~Allanach, P.~S.~B.~Dev, S.~A.~Renner and K.~Sakurai,
  arXiv:1512.07645;\\
  R. Ding, L. Huang, T. Li and B. Zhu, arXiv:1512.06560.\vspace*{-1mm} 

\bibitem{NMSSM} U. Ellwanger and C. Hugonie, arXiv:1602.03344;\\ 
F.~Domingo, S.~Heinemeyer, J.~S.~Kim and K. Rolbiecki, arXiv:1602.07691;\\ 
M.~Badziak, M.~Olechowski, S.~Pokorski and K.~Sakurai, arXiv:1603.02203;\\
P.~Baratella, J.~Elias-Miro, J.~Penedo and A.~Romanino,  arXiv:1603.05682.\vspace*{-1mm}

\bibitem{stop1} K. Hikasa and M. Kobayashi, Phys.Rev. D36 (1987) 724; \\ 
C. Boehm, A. Djouadi and Y. Mambrini, Phys. Rev. D61 (2000) 095006.\vspace*{-1mm}  

\bibitem{Zll-ATLAS} ATLAS Collaboration, Eur. Phys. J. C75 (2015) no.7, 
318.\vspace*{-1mm}  

\bibitem{stop2} C. Boehm, A. Djouadi and M. Drees, Phys.~Rev.~D62 (2000) 035012;\\
J.R. Ellis, K.A. Olive and Y. Santoso, Astropart.~Phys.~18 (2003) 395.\vspace*{-1mm} 

\bibitem{progress} A. Bharucha, A. Djouadi and A. Pilaftsis, work in progress.\vspace*{-1mm}   
     
\end{small}

\end{thebibliography}
\end{document}